\documentclass[11pt]{article}

\usepackage{latexsym,amssymb,psfig}

\newtheorem{defi}{Definition}
\newtheorem{theo}{Theorem}

\newtheorem{cor}[theo]{Corollary}

\newtheorem{con}{Conjecture}

\def\D{{D}}

\def\half{\frac{1}{2}}
\def\ket#1{|\,#1\,\rangle}

\newcommand{\textcirc}{% circle of reasonable size 
  \setlength{\unitlength}{1.5ex}%
  \begin{picture}(1.1,1)(-.55,-.55)%
    \put(0,0){\circle{1.1}}%
  \end{picture}}
\newcommand{\defend}{\hspace*{\fill} $\textcirc$}
\newcommand{\exend}{\hspace*{\fill} $\diamondsuit$}
\newcommand{\Tr}{{\rm Tr}}
\newcommand{\HA}{{\cal H}_{{A}}}
\newcommand{\HB}{{\cal H}_{{B}}}
\newcommand{\HE}{{\cal H}_{{E}}}
\newcommand{\oz}{\overline{z}}
\newcommand{\prob}{{\rm Prob\,}}
\newcommand{\cd}{{D}}

\newcommand{\noi}{\noindent}
\newcommand{\cancel}[1]{}
\newcommand{\canpro}[1]{}

\newcommand{\be}{\beta}
\newcommand{\ee}{\end{equation}}
\newcommand{\bea}{\begin{eqnarray}}
\newcommand{\beq}{\begin{equation}}
\newcommand{\eea}{\end{eqnarray}}
\newcommand{\beas}{\begin{eqnarray*}}
\newcommand{\eeas}{\end{eqnarray*}}
\newcommand{\bece}{\begin{center}}
\newcommand{\ence}{\end{center}}
\newcommand{\beit}{\begin{itemize}}
\newcommand{\enit}{\end{itemize}}

\newcommand{\proof}{{\it Proof. }}

\newcommand{\peon}{\hspace*{\fill} $\Box$\\}

\newcommand{\z}{{\bf Z}}

\newcommand{\al}{\alpha}

\newcommand{\ep}{\varepsilon}
\newcommand{\de}{\delta}

\newcommand{\ga}{\gamma}

\newcommand{\op}{\oplus}

\newcommand{\OX}{\overline{X}}

\newcommand{\OY}{\overline{Y}}
\newcommand{\OZ}{\overline{Z}}

\newcommand{\cx}{{\cal X}}

\newcommand{\cy}{{\cal Y}}

\newcommand{\CX}{{\cal X}}
\newcommand{\CY}{{\cal Y}}
\newcommand{\CZ}{{\cal Z}}

\newcommand{\ida}{I(X;Y\hspace{-1.3mm}\downarrow\hspace{-0.8mm} Z)}

\def\half{\frac{1}{2}}
 \def\<{\langle}
 \def\>{\rangle}
 
 \def\down{\downarrow}

 \def\opone{\leavevmode\hbox{\small1\kern-3.8pt\normalsize1}}
 \def\H{{\cal H}}
 \def\A{{\cal A}}
 \def\B{{\cal B}}
 \def\E{{\cal E}}
 \def\z{{\bar z}}
 \newcommand{\complex}{{\kern .1em {\raise .47ex\hbox {$\scriptscriptstyle
 |$}}\kern -.4em {\rm C}}}
 \newcommand{\real}{{{\rm I} \kern -.19em {\rm R}}}
\newcommand{\eeq}{\end{equation}}
 \newcommand{\beqa}{\begin{eqnarray}}
 \newcommand{\eeqa}{\end{eqnarray}}

\begin{document}

\title{
\bf Linking Classical and Quantum Key Agreement:\\
 Is There  ``Bound  Information''?}

\author{N. Gisin$^{(1)}$ and S. Wolf$^{(2)}$\\
\protect\small\em{(1) Group of Applied Physics, University of Geneva, 1211 Geneva,
Switzerland}\\
\protect\small\em{(2) Dept. of Computer Science, Swiss Federal Institute of Technology
(ETH Z\"urich),
ETH Zentrum, 
8092 Zurich, Switzerland}
}
\date{}
\maketitle

\begin{abstract}
\noi
After carrying out a protocol for quantum key agreement over a noisy quantum
channel, the parties Alice and Bob must process the raw key in order to 
end up with identical keys
about which 
the adversary has virtually no 
 information. 
In principle, both classical and quantum protocols can be used for this 
processing. It is a natural question which type of protocols is more 
powerful. We show that in many cases, for instance when 
the  4-state or the 6-state protocol is used, the limits of tolerable 
noise are identical for classical and quantum protocols. 
More precisely, we prove for general states   but 
under the assumption of incoherent eavesdropping that
Alice and Bob share some  so-called intrinsic  information
in their classical random variables,   resulting from optimal measurements,
if and only if the parties' quantum systems are entangled.
In addition, we provide evidence which strongly suggests that the 
potentials
of classical and of quantum protocols are equal in every situation.
It is an  important consequence of these parallels that
many techniques and results from quantum information theory
directly apply to
problems
in classical information theory, and vice versa.
For instance, it was previously believed 
that two parties can carry out unconditionally secure 
key agreement as long as they share some intrinsic 
information in the adversary's view.
The analysis of  this purely classical  problem from the 
quantum information-theoretic viewpoint shows
that this is
true in the binary case, but
false in general.
More explicitly,  
bound entanglement, i.e.,  entanglement that cannot be purified by any quantum 
protocol, 
has a classical counterpart. This ``bound intrinsic information'' cannot be 
distilled to a secret key by any classical  protocol. 
 As another application we propose a measure for 
entanglement based on classical information-theoretic quantities.
\end{abstract}

\section{Introduction}

In modern cryptography there are mainly two security paradigms, namely 
computational and information-theoretic security. The latter is sometimes also 
called unconditional security. Computational security is based on the 
assumed hardness of certain computational problems (e.g., the integer-factoring
or  discrete-logarithm problems). However, since a computationally 
sufficiently powerful adversary can solve any computational
 problem, hence break any 
such system, and because no useful general lower bounds are known in complexity 
theory, computational
 security is always conditional and, in addition to this, in danger by  progress
in the theory of efficient algorithms as well as in hardware engineering 
(e.g., quantum computing). 
Information-theoretic security on the other hand is based on probability 
theory and on the fact that an adversary's information is limited.
Such a limitation can for instance come from noise in communication 
channels or from the laws of quantum mechanics. 

Many different settings in the classical noisy-channel model have been 
described and analyzed, such as Wyner's wire-tap channel~\cite{wyner75},
Csisz\'ar and K\"orner's broadcast channel~\cite{csikor78}, or 
Maurer's  key agreement from joint randomness~\cite{ka},\,
 \cite{ittrans}.

Quantum cryptography on the other hand lies in the intersection of two of the 
major scientific achievements of the 20th century, namely quantum physics and 
information theory. Various protocols for so-called quantum
key agreement have been proposed  (e.g., \cite{BB84},\, \cite{ekert}), 
and
the possibility and impossibility of 
purification in different settings has been 
studied by many authors.

The goal of this paper is to derive parallels between classical and 
quantum key agreement
and thus to show that the two paradigms are 
more closely related  than previously recognized.
These connections allow for  investigating questions and  solving
open problems of purely classical information theory with quantum-mechanic
methods. 
One of the consequences is that, in contrast to what was 
previously believed, there exists a classical counterpart to so-called
{\em bound entanglement\/} (i.e., entanglement that cannot be purified by any quantum
protocol), namely intrinsic  information shared by Alice and Bob 
which they cannot use for generating a secret key by any classical protocol.

The outline of the paper is as follows. In Section~\ref{models} we introduce 
the classical (Section~\ref{classical}) and quantum (Section~\ref{quantum})
models 
of information-theoretic key agreement
and the crucial  concepts and quantities, such as  secret-key 
rate and intrinsic information on one side, and measurements,
entanglement, and quantum
privacy amplification on the other.
In Section~\ref{linking}, we show the mentioned links between these two 
models, more precisely, between entanglement and intrinsic information
(Section~\ref{linkingone}) as well as between quantum purification
and the secret-key rate (Section~\ref{prot}).
We illustrate the statements and their consequences with a number of examples
(Sections~\ref{exampleseins} and~\ref{exampleszwei} and Appendix~B). 
In Section~\ref{secbii} we define and characterize the classical 
counterpart of bound entanglement, called bound intrinsic information. 
Finally we show that 
not only   
problems  in classical information theory can be 
addressed by quantum-mechanical methods, but that the inverse is also  true:
In Section~\ref{linkingmeasure}
we propose a new measure for entanglement  
based on the intrinsic information measure.

\section{Models of Information-Theoretically Secure Key Agreement}
\label{models}

  \subsection{Key Agreement from Classical Information: 
              Intrinsic Information and  Secret-Key Rate}
\label{classical}

In this section we describe Maurer's general model of classical key agreement 
by public discussion from common information~\cite{ka}. 
In this setting, two parties 
Alice and Bob who are willing to generate a secret key have access to repeated
independent realizations of (classical) random variables $X$ and $Y$, 
respectively, whereas an adversary Eve learns the outcomes of a
random variable $Z$. Let $P_{XYZ}$ be the joint distribution of the three
random variables. In addition, Alice and Bob are 
connected by a noiseless and  authentic but otherwise completely insecure channel
(see Figure~\ref{maumod} in Appendix~A).
In this situation, the secret-key rate $S(X;Y||Z)$ has been defined as the
maximal rate at which Alice and Bob can generate a secret key that is equal
for Alice and Bob with overwhelming probability and about which Eve has  
only a negligible amount of (Shannon) information.
For a detailed discussion of the general scenario and the secret-key
rate as well as for various bounds on $S(X;Y||Z)$,
see~\cite{ka},\, \cite{strong},\, \cite{ittrans}.

Bound (\ref{lbo}) implies that if Bob's random variable
$Y$ provides more  information about Alice's $X$ than Eve's $Z$
does (or vice versa), 
then this advantage can be exploited for generating a secret key:
\beq\label{lbo}
S(X;Y||Z)\geq \max\, \{I(X;Y)-I(X;Z)\, ,\, I(Y;X)-I(Y;Z)\}\ .
\ee
This is a consequence of  a result by Csisz\'ar and 
K\"orner~\cite{csikor78}.
It is somewhat surprising 
that this bound is not tight, in particular, 
that secret-key agreement can even be possible 
when the right-hand side of (\ref{lbo}) vanishes or is negative. However,
the positivity of the expression on the right-hand side
of (\ref{lbo}) is a necessary and sufficient condition for the possibility of 
secret-key agreement by {\em one-way communication}: whenever Alice and 
Bob start in a disadvantageous situation with respect  to Eve,  {\em feedback\/} 
is necessary. The corresponding initial phase of the key-agreement protocol is
then often called {\em advantage distillation}.

The following upper bound on $S(X;Y||Z)$ 
is a generalization of Shannon's well-known impracticality theorem~\cite{shannon} 
and quantifies the intuitive fact that
no information-theoretically secure key agreement is possible when  Bob's information
is independent from  Alice's random variable, given Eve's information:
$
S(X;Y||Z)\leq I(X;Y|Z).
$
However, this bound is not tight. Because it is a possible strategy 
of the adversary Eve to process $Z$, i.e., to send $Z$ over some channel
characterized by $P_{\OZ|Z}$, we have for such a  new random variable $\OZ$
that
$
S(X;Y||Z)\leq I(X;Y|\OZ),
$
and hence 
\beq\label{ubo}
S(X;Y||Z)\leq\min\, _{P_{\OZ|Z}}\{I(X;Y|\OZ)\}=:\ida
\ee
holds.
The quantity $\ida$ has been called the {\em intrinsic conditional  information 
between $X$ and $Y$  given $Z$}~\cite{ittrans}. It was 
conjectured, and  evidence 
supporting this belief was given,
 that  $S(X;Y||Z)>0$ holds if $\ida>0$ does~\cite{ittrans}.
Some of the results below  
strongly suggest
that this is 
true if one of the random variables $X$ and $Y$ is binary and the other
one at most ternary, but false in general.

  \subsection{Quantum Key Agreement: Measurements, Entanglement, Purification}
\label{quantum}

We assume that the reader is familiar with the basic quantum-theoretic concepts
and notations. For an introduction, see for example~\cite{Peresbook}.

In the context of quantum key agreement, the classical scenario
$P_{XYZ}$ is replaced by a quantum state vector 
$\Psi\in\HA\otimes\HB\otimes\HE$\footnote{We consider pure states, since it
is natural to assume that Eve controls all the environment outside Alice
and Bob's systems.}, where $\HA$, $\HB$,
 and $\HE$ are Hilbert
spaces describing the systems in Alice's, Bob's, and Eve's hands, respectively.
Then, measuring this quantum state by the three parties leads to a classical 
probability distribution. In the following, we assume that Eve 
is free  
to carry out so-called {\em generalized measurements\/} (POVMs)~\cite{Peresbook}. In other words,
the set $\{|z\rangle\}$ will not be  assumed to be an orthonormal
 basis, but any set generating the Hilbert space
$\HE$ and satisfying the condition
$\sum_z|z\rangle\langle z|=\opone_{\HE}$. 
Then, if  the three parties carry out measurements in certain bases\footnote{We 
 assume all bases to be orthonormal.} $\{|x\rangle\}$ and
$\{|y\rangle\}$, and in the set $\{|z\rangle\}$, respectively, 
they end up with the classical scenario $P_{XYZ}=|\langle x,y,z|\Psi\rangle|^2$. 
Since this distribution depends on the chosen bases and set, 
 a given
quantum state $\Psi$ does not uniquely determine a classical scenario: some 
measurements may lead to scenarios useful for Alice and Bob, whereas for Eve, 
some others may (see Appendix~B).

The analog of Alice and Bob's marginal distribution $P_{XY}$
is the partial state 
$\rho_{AB}$,
obtained by tracing over Eve's Hilbert space $\HE$. 
More precisely, let 
$\Psi=\sum_{xyz}{c_{xyz}|x,y,z\rangle}$,
where $|x,y,z\rangle$ is short for $|x\rangle \otimes |y\rangle \otimes|z\rangle$.
We can write $\Psi=\sum_z{\sqrt{P_Z(z)}\psi_z\otimes |z\rangle}$, where 
$P_Z$ denotes Eve's marginal distribution of $P_{XYZ}$.
Then
$
\rho_{AB}=\Tr_{\HE}(P_{\Psi}):=\sum_z{P_Z(z)P_{\psi_z}}
$,
where $P_{\psi_z}$ is 
the projector to the state vector $\psi_z$. 

An important property is that  $\rho_{AB}$ is  pure 
($\rho_{AB}^2=\rho_{AB}$)
if and only if the global state $\Psi$ factorizes, i.e., 
$\Psi=\psi_{AB}\otimes\psi_E$,
where $\psi_{AB}\in\HA\otimes\HB$ and $\psi_E\in\HE$. 
In this case Alice and Bob
are independent of Eve: Eve cannot obtain any information on Alice's and Bob's
states by
measuring her system.

 After a measurement,
Alice and Bob obtain a classical distribution $P_{XY}$. However, in order to obtain a well-defined
classical scenario one has to assume that also Eve performs a  measurement, i.e., that
Eve treats her information on the classical level. Indeed, only then a classical
distribution $P_{XYZ}$ is defined. But considering that in practice all $P_{XYZ}$ result
from some physical process, the assumption that Eve performs the measurement one would
like her to perform is 
not founded on basic principles\footnote{One could argue that if the system in Eve's hand is
classical, then she has no choice for her measurement. But ultimately all systems are
quantum mechanical and the apparent lack of choice might  purely be a matter of 
technology.}. For example, Eve's measurement 
could be done later and 
depend on the
public discussion between Alice and Bob. Consequently, the common approach which
starts from $P_{XYZ}$ to prove the security of a key agreement protocol hides an 
assumption
about Eve's measurement.
As we shall see, avoiding this hidden assumption 
and staying in the quantum regime 
can actually simplify the analysis of
the scenario.

When Alice and Bob share many independent 
systems\footnote{Here we do not consider the possibility that
Eve coherently processes several of her systems. This corresponds to the 
assumption in the classical scenario that repeated realizations of $X$, $Y$, and 
$Z$  are independent of each other.} 
$\rho_{AB}$,
there are basically two possibilities for 
generating a secret key.
Either they first measure
their systems and then run a classical protocol (process classical information)
secure against all measurements Eve could possibly perform
(i.e., against all possible distributions $P_{XYZ}$ that can result after 
Eve's measurement). Or they first run
a quantum protocol (i.e., process the information in the quantum domain) and then
perform their measurements.
The idea of quantum protocols is to process the systems in state $\rho_{AB}$ and to
produce fewer systems  in a pure state (i.e., to {\em purify\/} $\rho_{AB}$), thus to
eliminate Eve from the scenario. Moreover, the pure state Alice and Bob end up
with should be maximally entangled 
(i.e., even for some different and incompatible measurements, Alice's
and Bob's results are perfectly correlated). 
Finally, Alice and Bob measure their maximally entangled 
systems and establish a secret key. This
way of obtaining a  key directly from a quantum state $\Psi$, without any
error correction nor classical privacy amplification, is called {\em quantum privacy
amplification\/}\footnote{
The term ``quantum privacy amplification'' is somewhat unfortunate since
it does not correspond to classical privacy amplification, but includes advantage
distillation and error correction.} (QPA for short)~\cite{QPA},\, \cite{benal}.
Note that the procedure described in~\cite{QPA} and~\cite{benal}
 guarantees that Eve's {\em relative\/} information 
       (relative to the key length) is arbitrarily small, but not 
that her 
       {\em absolute\/} information is negligible. The analog of this problem 
       in the classical case is
       discussed in~\cite{strong}.

The precise conditions under which a general state $\rho_{AB}$ can be purified are
not known. However, the two following conditions are necessary. First, the state must
be {\em entangled\/} or, equivalently, not {\em separable}. A state $\rho_{AB}$ is separable 
 if and only if it can be written as a mixture of product states, i.e., 
$
\rho_{AB}=\sum_j p_j \rho_{Aj}\otimes\rho_{Bj}
$.
Separable states can be generated by purely classical communication, hence it 
follows from bound~(\ref{ubo}) that 
 entanglement is a necessary condition. The second 
condition is more subtle: The matrix $\rho_{AB}^t$ obtained from $\rho_{AB}$
by {\em partial transposition\/}
must have at least one negative eigenvalue~\cite{HorodeckiPartialTransp},
\, \cite{Horodecki}. 
The partial transposition of the density matrix $\rho_{AB}$ is defined as
$
(\rho_{AB})_{i,j;\mu,\nu}^t:=(\rho_{AB})_{i,\nu;\mu,j}
$,
where the indices $i$ and $\mu$ [$j$ and $\nu$] 
run through a basis of $\HA$ [$\HB$].
Note that this definition is base-dependent. However, the {\em eigenvalues\/} 
of $\rho_{AB}^t$ are not~\cite{Peres}. The second of these conditions implies
the first one: negative (i.e., at least one eigenvalue is negative) 
partial transposition implies
entanglement.

In the binary case ($\HA$ and $\HB$ both have dimension two), the above two
conditions are equivalent and  sufficient for the possibility of quantum key agreement: 
all  entangled binary  states can be
purified. 
The same even
holds if one Hilbert space is of dimension 2 and the other one of dimension 3.
However, for larger dimensions there are
examples showing that these conditions are not equivalent: There are entangled states whose
partial transpose has no negative eigenvalue, hence cannot be 
purified~\cite{HorodeckiPartialTransp}. 
Such states are called {\em bound entangled}, in contrast to {\em free entangled\/}
states,
 which can be
purified.
Moreover, it is believed that 
there even exist entangled states which cannot be purified although they have negative 
partial transposition~\cite{benpt}.

\section{Linking Classical and Quantum Key Agreement}
\label{linking}

In this section we derive a close connection 
 between the possibilities offered by classical
and quantum protocols for key agreement. The intuition is as follows. 
As described in Section~\ref{quantum},
there is a very natural connection between quantum states $\Psi$ 
 and 
classical distributions $P_{XYZ}$ which 
can be thought of as arising from $\Psi$
by measuring in a certain basis, e.g., the standard basis\footnote{A priori, there is no
privileged basis. However,  physicists often write states like $\rho_{AB}$ in a
basis which seems to be more natural than others. 
We refer to this   as the standard
basis. Somewhat surprisingly, this basis is generally easy to identify,
though not precisely defined. One could  characterize the
standard basis as the basis for which as many coefficients as possible of $\Psi$ are
real and positive.
We usually represent quantum states with respect to the standard basis.}. 
(Note however that the connection is not unique  even for fixed bases: 
For a given distribution $P_{XYZ}$, there are 
many states $\Psi$ leading to $P_{XYZ}$ by carrying out measurements.) 
When given a state $\Psi$ between three parties 
Alice, Bob, and Eve, 
and if $\rho_{AB}$ denotes the resulting mixed state  after tracing out Eve,
then the corresponding classical distribution $P_{XYZ}$
has positive intrinsic information if and only if 
$\rho_{AB}$ is entangled. However, 
this correspondence clearly depends on the measurement bases used by Alice, Bob, and Eve.
If for instance $\rho_{AB}$ is entangled, but Alice and Bob do very unclever
measurements, then the intrinsic information may vanish (see Example~7 in Appendix~B). 
If on the other hand $\rho_{AB}$ is 
separable, Eve may do such bad measurements that the intrinsic information
becomes positive, despite the fact that $\rho_{AB}$ could have
been established by public discussion without any prior correlation (see Example~6 
in Appendix~B). 
Consequently, the correspondence between
intrinsic information and entanglement must involve some optimization over all
possible measurements on all sides.

A similar  correspondence on the protocol level is supported 
by many examples, but not  rigorously proven:
The distribution $P_{XYZ}$ allows for classical key agreement if and 
only if quantum key agreement is possible starting from the state $\rho_{AB}$.

We show how these parallels allow for addressing problems
of purely classical information-theoretic nature 
with the methods of quantum information theory, and vice versa.

\subsection{Entanglement and Intrinsic Information}
\label{linkingone}

Let us first establish the connection between intrinsic information
and entanglement.   Theorem~\ref{theoeins} states  that if 
 $\rho_{AB}$ is separable, then Eve can ``force'' the  information
between
   Alice's and Bob's classical random variables (given Eve's classical random variable) 
to be zero (whatever strategy Alice and Bob use). 
   In particular, Eve can prevent classical key agreement.

\begin{theo}\label{theoeins}
Let $\Psi\in\HA\otimes\HB\otimes\HE$ and $\rho_{AB}=\Tr_{\HE}(P_{\Psi})$.
If $\rho_{AB}$ is separable,
then there exists a generating set $\{|z\rangle\} $ of $\H_\E$  such that
for all bases $\{|x\rangle\} $ and $\{|y\rangle\} $ of $\H_\A$ and $\H_\B$,
respectively,
$I(X;Y|Z)=0$
holds for 
 $P_{XYZ}(x,y,z):=|\langle x,y,z|\Psi\rangle |^2$.
\end{theo}

\noi
\proof
If $\rho_{AB}$ is separable, then there exist vectors $|\alpha_z\rangle$ and
$|\beta_z\rangle$
such that
$
\rho_{AB}=\sum_{z=1}^{n_z} p_z
P_{\alpha_z}\otimes P_{\beta_z}
$,
where $P_{\alpha_z}$ denotes the one-dimensional projector onto the subspace spanned
by $|\alpha_z\rangle$.

Let us first assume that $n_z\le \dim\HE$.
Then there exists a basis $\{|z\rangle\}$ of $\HE$ such that 
$
\Psi=\sum_z \sqrt{p_z}\, |\alpha_z,\beta_z,z\rangle 
$
holds~\cite{qp9609013},\, 
\cite{gisin89},\, \cite{hjw93}.

If $n_z> \dim\HE$, then Eve can add an auxiliary system $\H_{aux}$
to hers (usually called an {\em ancilla\/}) and we have
$
\Psi\otimes|\gamma_0\rangle=\sum_z \sqrt{p_z}\,
|\alpha_z,\beta_z,\gamma_z\rangle
$,
where $|\gamma_0\rangle\in\H_{aux}$ is the state of Eve's auxiliary
system, and 
$\{|\gamma_z\rangle\}$ is a basis of $\HE\otimes\H_{aux}$.
We define the (not necessarily orthonormalized) vectors $|z\rangle$ by 
$|z,\gamma_0\rangle=\opone_{\HE}\otimes P_{\gamma_0}|\gamma_z\rangle$.
These vectors determine a generalized measurement with positive operators
$O_z=|z\rangle\langle z|$.
Since $\sum_zO_z\otimes
P_{\gamma_0}=\sum_z|z,\gamma_0\rangle\langle z,\gamma_0|
=\sum_z\opone_{\HE}\otimes
P_{\gamma_0}|\gamma_z\rangle\langle|\gamma_z| \opone_{\HE}\otimes
P_{\gamma_0}
=\opone_{\HE}\otimes P_{\gamma_0}$, the $O_z$ satisfy
$\sum_zO_z=\opone_{\HE}$, as 
they should in order to define a generalized measurement \cite{Peresbook}.
Note that the first case ($n_z\le \dim\HE$) is a special case of the
second one, with $|\gamma_z\rangle=|z,\gamma_0\rangle$.
If Eve now performs the measurement, then we have $P_{XYZ}(x,y,z)=|\langle
x,y,z|\Psi\rangle |^2
=|\langle x,y,\gamma_z|\Psi,\gamma_0\rangle |^2$, and
\[
P_{XY|Z}(x,y,z)=|\langle x,y|\alpha_z,\beta_z\rangle |^2
=|\langle x|\alpha_z\rangle |^2\, |\langle y|\beta_z\rangle
|^2=P_{X|Z}(x,z)P_{Y|Z}
(y,z)
\]
holds for all $|z\rangle$ and for all
$|x,y\rangle \in\H_\A\otimes\H_\B$. Consequently, $I(X;Y|Z)=0$.
\peon 
\ 

Theorem~\ref{theozwei} states that 
 if $\rho_{AB}$ is  entangled,
   then Eve {\em cannot\/} force the intrinsic information to be zero: Whatever
   she does (i.e., whatever generalized measurements she carries out), there 
   is something Alice and Bob can do such that the intrinsic information
   is positive. Note that this does {\em not}, a priori, imply that secret-key
   agreement is possible in every case. Indeed, we will provide 
   evidence for the fact that this implication does generally {\em not\/}
   hold.    

\begin{theo}\label{theozwei}
Let $\Psi\in\HA\otimes\HB\otimes\HE$ and $\rho_{AB}=\Tr_{\HE}(P_{\Psi})$.
If $\rho_{AB}$ is entangled,
then for all generating sets $\{|z\rangle\} $ of $\H_\E$, 
there are bases $\{|x\rangle\} $ and $\{|y\rangle\} $ of $\H_\A$ and $\H_\B$, 
respectively,
such that
$I(X;Y\down Z)>0$
holds for
 $P_{XYZ}(x,y,z):=|\langle x,y,z|\Psi\rangle |^2$.
\end{theo}

\noi
\proof
We prove this by contradiction.
Assume 
that there exists a generating set $\{|z\rangle\}$ of $\HE$ such that for 
all bases $\{|x\rangle\}$ of $\HA$ and $\{|y\rangle\}$ of $\HB$,
$I(X;Y\down Z)=0$ holds for the resulting distribution. 
For such a distribution,  there exists a channel, characterized by
$P_{\OZ|Z}$,
such that $I(X;Y|\OZ)=0$ holds, i.e., 
\beq
P_{XY|\OZ}(x,y,\overline{z})=P_{X|\OZ}(x,\overline{z})P_{Y|\OZ}(y,\overline{z})\ .
\label{pbar}
\eeq
Let now 
$
\rho_{\overline{z}}
:=(1/p_{\overline{z}})\sum_z p_z P_{\OZ|Z}(\overline{z},z)P_{\psi_z}
$
with $p_z=P_Z(z)$ and
$
p_{\overline{z}}=\sum_z P_{\OZ|Z}(\overline{z},z)p_z,
$
and where $\psi_z$ is the state of Alice's and Bob's system conditioned on Eve's
result $z$:
$\Psi\otimes|\gamma_0\rangle=\sum_z\psi_z\otimes|\gamma_z\rangle$
(see the proof of Theorem~1).

 From (\ref{pbar}) we can conclude
$
\Tr(P_x\otimes P_y\rho_{\overline{z}})=\Tr(P_x\otimes \opone\rho_{\bar
z})\, \Tr(\opone\otimes P_y\rho_{\overline{z}})
$
for all one-dimensional
 projectors $P_x$ and $P_y$ acting in $\H_\A$ and $\H_\B$,
respectively.
Consequently, the states $\rho_{\overline{z}}$ are products, i.e., 
$
\rho_{\overline{z}}=\rho_{\alpha_{\overline{z}}}\otimes\rho_{\beta_{\overline{z}}},
$
and $\rho_{AB}=\sum_\z p_\z \rho_\z$ is separable.
\peon \

Theorem~\ref{theozwei} can be formulated in a more 
positive way. Let us first 
introduce the concept of a set of bases $\{|x^j\rangle,|y^j\rangle\} $,
where the $j$
   label the different bases, as they are used in the 4-state (2 bases) and
   the 6-state (3 bases) protocols~\cite{BB84},\, \cite{Dagmar6state},\, \cite{bg}. 
Then if $\rho_{AB}$ is entangled there exists a set
$\{|x^j\rangle,|y^j\rangle\}$ of $N$
bases such that for all generalized measurements $\{|z\rangle\}$,
$
I(X;Y\down [Z,j])>0
$
holds.
The idea is that Alice and Bob randomly choose a basis and, after the
transmission, publicly restrict to the (possibly few) cases where they
happen to have
chosen the same basis. Hence Eve knows $j$, and one has
$
I(X;Y\down [Z,j])=(1/N)\sum_{j=1}^N I(X^j;Y^j\down Z)\ .
$
If the set of bases is large enough, then for all $\{|z\rangle\}$
 there is a basis with
positive intrinsic information, 
hence the mean is also positive. 
Clearly, this result is 
stronger if the set of bases is small. 
Nothing is proven about 
the achievable size of such sets of bases, but it is conceivable that
$\max\{\dim\HA,\dim\HB\}$
 bases
are always sufficient.

\begin{cor}\label{coreinszwei}
Let $\Psi\in\HA\otimes\HB\otimes\HE$ and $\rho_{AB}=\Tr_{\HE}(P_{\Psi})$.
Then the following statements
are equivalent:
\\
\quad
{\it (i)}
$\rho_{AB}$ is entangled,
\\
\quad
{\it (ii)}
for all generating sets $\{|z\rangle\}$ of $\HE$, there exist bases
$\{|x\rangle\}$ of $\HA$ and 
$\{|y\rangle\}$ of  $\HB$ such that the 
distribution  $P_{XYZ}(x,y,z):=|\langle x,y,z|\Psi\rangle|^2$ satisfies 
$\ida>0$,
\\
\quad
{\it (ii)}
for all generating sets  $\{|z\rangle\}$ of $\HE$, there exist bases
$\{|x\rangle\}$ of $\HA$
and $\{|y\rangle\}$ of  $\HB$ such that the 
distribution  $P_{XYZ}(x,y,z):=|\langle x,y,z|\Psi\rangle|^2$ satisfies 
$I(X;Y|Z)>0$.
\end{cor}

A first consequence of the fact that Corollary~\ref{coreinszwei} often holds
with respect to the standard bases (see below) is that it yields, at least 
in the binary case, a criterion for $\ida>0$ that is efficiently verifiable 
since it is based on the positivity of the eigenvalues of a $4\times 4$ matrix
(see also Example~5). 
Previously, the quantity $\ida$ has been considered to be quite hard to handle.

\subsection{Examples I}
\label{exampleseins}

The following examples illustrate the correspondence established in 
Section~\ref{linkingone}. 
They show in particular that very
often (Examples 1, 2, and
3), but not always (Examples 6 and 7 in Appendix~B), the direct connection between 
entanglement and positive intrinsic information holds with respect 
to the standard bases (i.e., the bases physicists use by commodity
and intuition).
\ \\

\noi
{\it Example 1.}
Let us consider the  so-called 4-state  protocol
of~\cite{BB84}.
The analysis of the  6-state protocol~\cite{bg} is analogous and leads to similar
results~\cite{giswol99}. 
We compare  the possibility of 
quantum and classical key agreement given the 
quantum state and the corresponding classical distribution, respectively, arising 
from this protocol. The conclusion 
is, under the assumption of incoherent eavesdropping, that key 
agreement in one setting is possible if and only if this is true 
also for the other. 

After carrying out the 4-state protocol, and under the assumption of 
optimal  eavesdropping (in terms of Shannon information), the resulting 
quantum state is~\cite{FGGNP}
\[
\Psi=\sqrt{F/2}|0,0\rangle\otimes \xi_{00}+\sqrt{D/2}|0,1\rangle\otimes  \xi_{01}+\sqrt{D/2}|1,0\rangle\otimes  \xi_{10}+
\sqrt{F/2}|1,1\rangle\otimes  \xi_{11}\, \in{\bf C}^2\otimes {\bf C}^2\otimes {\bf C}^4\ ,
\]
where $D$ (the {\em disturbance\/}) is the probability that $X\ne Y$ holds if
 $X$ and $Y$ are the classical random variables of Alice and Bob, respectively,
where $F=1-D$ (the {\em fidelity\/}),
and where the $\xi_{ij}$ satisfy $\langle \xi_{00}|\xi_{11}\rangle=
\langle \xi_{01}|\xi_{10}\rangle=1-2D$ and $\langle \xi_{ii}|\xi_{ij}\rangle=0$
for all $i\ne j$. Then the state $\rho_{AB}$ is (in the  basis
$\{\ket{00}$,$\ket{01}$,$\ket{10}$,$\ket{11}\}$)
{\small
\[
\rho_{AB}=\half\pmatrix{\D & 0 & 0 & -\D(1-2D) \cr 0 & 1-D & -(1-D)(1-2D) & 0\cr 
0 & -(1-D)(1-2D) & 1-D & 0\cr -\D(1-2D) & 0 & 0 & \D}\ ,
\]
}and its partial transpose
{\small
\[
\rho_{AB}^t=\half\pmatrix{\D & 0 & 0 & -(1-D)(1-2D) \cr 0 & 1-D & -\D(1-2D) & 0\cr 
0 & -\D(1-2D) & 1-D & 0\cr -(1-D)(1-2D)  & 0 & 0 & \D}
\]
}has the eigenvalues $(1/2)(D\pm(1-D)(1-2D))$ and $(1/2)((1-D)\pm D(1-2D))$, which are 
all non-negative (i.e., $\rho_{AB}$ is separable) if
\beq\label{coeins}
D\geq 1-\frac{1}{\sqrt{2}}\ .
\eeq

From the classical viewpoint, the corresponding distributions (arising
 from measuring the above quantum system in the standard bases) 
are as follows. 
First, $X$ and $Y$ are both symmetric bits with $\prob[X\ne Y]=\cd$. 
Eve's random variable $Z=[Z_1,Z_2]$ is composed of 2 bits $Z_1$ and $Z_2$,
where $Z_1=X\oplus Y$, i.e., $Z_1$ tells Eve whether Bob received the qubit
disturbed ($Z_1=1$) 
or not ($Z_1=0$) (this is a consequence of the fact that the $\xi_{ii}$ and
$\xi_{ij}$ ($i\ne j$) states
 generate orthogonal sub-spaces), and where
the probability that Eve's second bit indicates the correct value
of Bob's bit 
is
Prob$[Z_2=Y]=\delta=(1+\sqrt{1-\langle \xi_{00}|\xi_{11}\rangle^2})/2=1/2+\sqrt{D(1-D)}$. 
We now prove that for this distribution, 
the intrinsic information is zero 
if and only if 
\beq\label{bsbed}
\frac{\cd}{1-\cd}\geq 2\sqrt{(1-\de)\de}=1-2D
\ee
holds. 
We show that if the condition (\ref{bsbed})
 is 
satisfied, then $\ida=0$ holds. 
The inverse implication follows from the existence of a key-agreement 
protocol in all other cases (see Example~1 (cont'd) in Section~\ref{exampleszwei}).
If  (\ref{bsbed})  holds, we can construct a random variable $\OZ$,
that is generated by sending $Z$ over a channel characterized by
$P_{\OZ|Z}$,  for which  $I(X;Y|\OZ)=0$ holds. 
We can restrict ourselves to the case of equality  in (\ref{bsbed})
because Eve can always increase $\de$ by adding noise. 

Consider now the channel characterized by the following distribution
$P_{\OZ|Z}$ (where $\overline{{\cal Z}}=\{u,v\}$):
$
P_{\OZ|Z}(u,[0,0])  =  P_{\OZ|Z}(v,[0,1]) = 1$, 
$P_{\OZ|Z}(l,[1,0])  =  P_{\OZ|Z}(l,[1,1]) = 1/2
$
for $l\in\{u,v\}$.
(The channel $P_{\OZ|Z}$ is 
illustrated in Figure~\ref{channel} in Appendix~A.)
We show $I(X;Y|\OZ)={\rm E}_{\OZ}\, [I(X;Y|\OZ=\oz)]=0$, i.e., that 
$I(X;Y|\OZ=u)=0$ and $I(X;Y|\OZ=v)=0$ hold. By symmetry it is sufficient 
to show the first equality. For  $a_{ij}:=P_{XY\OZ}(i,j,u)$, we get
\[
a_{00}  =  (1-D)(1-\de)/2\, ,\ 
a_{11}  =  (1-D)\de/2\, ,\ 
a_{01}  =  a_{10} = 
(D(1-\de)/2+D\de/2)/2=D/4\, .
\]
From  equality  in (\ref{bsbed})  we conclude 
$
a_{00}a_{11}=a_{01}a_{10}
$,
which is equivalent to the fact that $X$ and $Y$ are independent, given
$\OZ=u$.

Finally, note that the conditions (\ref{coeins}) and (\ref{bsbed}) are 
equivalent for $D\in [0,1/2]$. This shows that the bounds of tolerable noise
are indeed exactly the same for the quantum and classical scenarios.
\exend
\\ \

\noi
{\it Example 2.}
We consider the bound entangled state presented in~\cite{HorodeckiPartialTransp}. This
example received quite a lot of attention by the quantum-information community
because it was the first known example of bound entanglement (i.e., entanglement without
the possibility of
quantum key agreement). We show that 
its classical counterpart seems to have similarly surprising properties.
 Let $0<a<1$ and
\[
\Psi=\sqrt{\frac{3a}{8a+1}}\, \psi\otimes|0\rangle  +
\sqrt{\frac{1}{8a+1}}\, \phi_a\otimes|1\rangle  + 
\sqrt{\frac{a}{8a+1}}\, (|122\rangle +|133\rangle +|214\rangle +|235\rangle +|326\rangle )\ ,
\]
where $\psi=(|11\rangle +|22\rangle +|33\rangle )/\sqrt{3}$ and 
$
\phi_a=\sqrt{(1+a)/(2)}\, |31\rangle +\sqrt{(1-a)/(2)}\, |33\rangle.
$
It has been shown in~\cite{HorodeckiPartialTransp} that the resulting state 
$\rho_{AB}$
is entangled.

The
corresponding classical
distribution is 
 as follows. The ranges are
$
\CX=\CY=\{1,2,3\}$ and $\CZ=\{0,1,2,3,4,5,6\}
$.
We write
$
(ijk)=P_{XYZ}(i,j,k)
$.
Then we have
$
(110)=(220)=(330)=(122)=(133)=(214)=(235)=(326)  =  2a/(16a+2)$,
$(311)  =  (1+a)/(16a+2)$, and 
$(331)  =  (1-a)/(16a+2)$.
We study the special case $a=1/2$. Consider the following representation
of the resulting distribution (to be normalized).
For instance,
the entry ``$(0)\ 1\ ,\ (1)\ 1/2$'' for $X=Y=3$  means 
$P_{XYZ}(3,3,0)=1/10$ (normalized),
$P_{XYZ}(3,3,1)=1/20$,
and  $P_{XYZ}(3,3,z)=0$ for all $z\not\in\{0,1\}$.
\\
\begin{center}
\begin{tabular}{|c||c|c|c|}
\hline
\ \ X & 1 & 2 & 3\\
Y (Z) &&&\\
\hline\hline
1 & (0)\ 1 & (4)\ 1 & (1)\ 3/2\\
\hline
2 & (2)\ 1 & (0)\ 1 & (6)\ 1\\
\hline
3 & (3)\ 1 & (5)\ 1 & (0)\ 1\\
&&& (1)\ 1/2\\
\hline
\end{tabular}
\end{center}

As we would expect, the  intrinsic information is positive 
in this scenario. This can be seen by contradiction as follows. 
Assume $\ida=0$. Hence there exists a discrete channel, characterized
by the conditional distribution $P_{\OZ|Z}$, such that $I(X;Y|\OZ)=0$ holds.
Let $\overline{\CZ}\subseteq {\bf N}$, and let
$
P_{\OZ|Z}(i,0)  =:  a_i$,
$P_{\OZ|Z}(i,1)  =:  x_i$,
$P_{\OZ|Z}(i,6)  =:  s_i$.
Then we must have 
$a_i,x_i,s_i\in [0,1]$ and 
$\sum_i{a_i}=\sum_i{x_i}=\sum_i{s_i}=1$.
Using  $I(X;Y|\OZ)=0$, we obtain the following distributions 
$P_{XY|\OZ=i}$ (to be normalized):
\\
\begin{center}
\begin{tabular}{|c||c|c|c|}
\hline
\ \ X & 1 & 2 & 3\\
Y  &&&\\
\hline\hline
1 & $a_i$ & $\frac{3a_ix_i}{2s_i}$ & $\frac{3x_i}{2}$\\
\hline
2 & $\frac{2a_is_i}{3x_i}$ & $a_i$ & $s_i$\\
\hline
3 & $\frac{2a_i(a_i+x_i/2)}{3x_i}$ & $\frac{a_i(a_i+x_i/2)}{s_i}$ & 
$a_i+\frac{x_i}{2}$\\
\hline
\end{tabular}
\end{center}
By comparing the $(2,3)$-entries of the two tables above, we obtain
\beq\label{drei}
1\geq\sum_i{\frac{a_i(a_i+x_i/2)}{s_i}}\ .
\ee

We now prove that  (\ref{drei}) 
implies $s_i\equiv a_i$ (i.e., $s_i =a_i$ for all $i$) and $x_i\equiv 0$. 
Clearly, this 
does not lead to a solution and is hence
a 
contradiction. For instance,
$
P_{XY|\OZ=i}(1,2)=2a_is_i/(3x_i)
$
is not even defined in this case if $a_i>0$.

It remains to show that  (\ref{drei})
implies $a_i\equiv s_i$ and $x_i\equiv 0$. We show that 
whenever $\sum_i{a_i}=\sum_i{s_i}=1$ and $a_i\not\equiv s_i$, then
$
\sum_i{a_i^2/s_i}>1\ .
$
First, note that $\sum_i{a_i^2/s_i}=\sum_i{a_i}=1$ for $a_i\equiv s_i$.
Let now $s_{i_1}\leq a_{i_1}$ and $s_{i_2}\geq a_{i_2}$. We show that 
$
a_{i_1}^2/s_{i_1}+a_{i_2}^2/s_{i_2}<a_{i_1}^2/(s_{i_1}-\ep)+a_{i_2}^2/(s_{i_2}+\ep)
$
holds for every $\ep>0$,
which obviously  implies  the above statement. It is  straightforward 
to see that this
 is equivalent to
$
a_{i_1}^2s_{i_2}(s_{i_2}+\ep)>a_{i_2}^2s_{i_1}(s_{i_1}-\ep),
$
and holds because of
$
a_{i_1}^2s_{i_2}(s_{i_2}+\ep)>a_{i_1}^2a_{i_2}^2$ and 
$
a_{i_2}^2s_{i_1}(s_{i_1}-\ep)<a_{i_1}^2a_{i_2}^2
$.
This concludes the proof of  $\ida>0$.
\exend
\ \\

As mentioned,
the interesting point about Example~2 is that the quantum state is
bound entangled, and that also classical key agreement seems 
impossible despite the fact that $\ida>0$ holds. This is a  contradiction 
 to a conjecture stated in~\cite{ittrans}. 
The classical translation of the bound entangled state leads to a classical 
distribution with very strange properties as well! (See 
Example~2 (cont'd) in Section~\ref{prot}).

In
Example~3, another  bound entangled state (first proposed in~\cite{dreih}) is discussed. 
The example  is particularly 
nice because, depending on the choice of the parameter $\al$, the quantum 
state can be made separable, bound entangled, and free entangled.
\\ \

\noi
{\it Example 3.}
We consider the following  distribution (to be normalized).
Let $2\leq \al \leq 5$.
\\
\begin{center}
\begin{tabular}{|c||c|c|c|}
\hline
\ \ X & 1 & 2 & 3\\
Y (Z) &&&\\
\hline\hline
1 & (0)\ 2 & (4)\ $5-\al$ & (3)\ $\al$\\
\hline
2 & (1)\ $\al$ & (0)\ 2 & (5)\ $5-\al$\\
\hline
3 & (6)\ $5-\al$ & (2)\ $\al$ & (0)\ 2\\
\hline
\end{tabular}
\end{center}
This distribution arises when measuring the following quantum state.
Let 
$
\psi:=(1/\sqrt{3})\, (|11\rangle+|22\rangle+|33\rangle).
$
Then 
\[
\Psi=\sqrt{\frac{2}{7}}\, \psi\otimes|0\rangle + 
\sqrt{\frac{a}{21}}\, (|12\rangle\otimes|1\rangle+|23\rangle\otimes|2\rangle
+|31\rangle\otimes|3\rangle) + 
\sqrt{\frac{5-a}{21}}\, (|21\rangle\otimes|4\rangle +|32\rangle\otimes|5\rangle
+|13\rangle\otimes|6\rangle),
\]
\[
\mbox{and\ \ \ \ \quad\ }
\rho_{AB}=\frac{2}{7}\, P_{\psi} + \frac{a}{21}\, (P_{12}+P_{23}+P_{31}) + \frac{5-a}{21}\, (P_{21}+P_{32}+P_{13})
\]
is separable if and only if $\al\in[2,3]$, bound entangled for $\al\in(3,4]$, and 
free entangled if $\al\in(4,5]$~\cite{dreih} (see Figure~\ref{qcw} in Appendix~A).

Let us consider the quantity $\ida$. First of all, it is clear that 
$\ida=0$ holds for $\al\in[2,3]$. The reason is that 
$
\al\geq 2$ and $5-\al\geq 2
$
together imply that Eve can ``mix'' her symbol $Z=0$ with the remaining 
symbols  in such a way that when given that $\OZ$ takes the ``mixed value,'' then
$XY$ is uniformly distributed; in particular, $X$ and $Y$ are 
independent. Moreover, it can be shown in analogy to 
Example 2 that $\ida>0$ holds for  $\al>3$.
\exend
\\ \

Examples~1, 2, and~3 suggest that the correspondence between separability
and entanglement on one side and vanishing and non-vanishing intrinsic 
information on the other always holds with respect to the standard bases
or even arbitrary bases.
We show in Appendix~B that this is 
not true in general. More precisely, Examples~6 and~7 demonstrate how Eve as well as
Alice and Bob can perform bad measurements. Hence the parallelity between the 
quantum and classical situation must be  as it is stated in Theorems~1 and~2.

  \subsection{A Classical Measure for Quantum Entanglement}
\label{linkingmeasure}

It is a
challenging problem
of theoretical quantum physics to find good measures for entanglement~\cite{qp9610044}.
Corollary~\ref{coreinszwei} above
   suggests the following measure,  which is based on
    classical information theory. 
\begin{defi}
{\rm
Let for a quantum state $\rho_{AB}$
\[
\mu(\rho_{AB}):= \min_{\{|z\rangle\}}\, (\max_{\{|x\rangle\},\{|y\rangle\}}\, (\ida))\ ,
\]
where the minimum is taken over all $\Psi=\sum_z\sqrt{p_z}\psi_z\otimes|z\rangle $
such that
$\rho_{AB}=\Tr_{\HE}(P_{\Psi})$ holds
and over all bases $\{|z\rangle\}$ of $\HE$, the maximum is over all bases $\{|x\rangle\}$
of $\HA$ and $\{|y\rangle\}$
of $\HB$, and where
$P_{XYZ}(x,y,z):=|\langle x,y,z|\Psi\rangle |^2$.
}
\defend
\end{defi}

Then $\mu$ has all the properties required from such a measure.
If $\rho_{AB}$ is pure, i.e., $\rho_{AB}=|\psi_{AB}\rangle\langle\psi_{AB}|$, 
then we have  in the Schmidt
basis (see for example~\cite{Peresbook}) 
$\psi_{AB}=\sum_j c_j |x_j,y_j\rangle $, and
$
\mu(\rho_{AB})=-\sum_j|c_j|^2\log_2(|c_j|^2)=
-\Tr(\rho_{AB}\log_2\rho_{AB})
$,
 as it should~\cite{qp9610044}.
It is obvious that $\mu$ is convex, i.e., 
$\mu(\lambda\rho_1+(1-\lambda)\rho_2)\le\lambda\mu(\rho_1)+(1-\lambda)\mu(
\rho_2)$.
\\ \

\noi
{\it Example~4 (based on Werner's states).}
Let
$
\Psi=\sqrt{\lambda}\, \psi^{(-)}\otimes|0\rangle  +
\sqrt{(1-\lambda)/4}\, |001+012+103+114\rangle
$,
where $\psi^{(-)}=|10-01\rangle /\sqrt{2}$, and $\rho_{AB}=\lambda P_{\psi^{(-)}} +
((1-\lambda)/4)\opone$.
It is well-known that $\rho_{AB}$ is separable if and only if 
 $\lambda\le 1/3$.
Then the classical distribution is $(010)=(100)=\lambda/2$ and 
$(001)=(012)=(103)=(114)=(1-\lambda)/4$.

If $\lambda\le1/3$, then consider the channel
$P_{\OZ|Z}( 0,0)=
P_{\OZ|Z}(  2,2)=
P_{\OZ|Z}(  3,3)=1\, ,\ 
P_{\OZ|Z}(  0,1)=
P_{\OZ|Z}(  0,4)=\xi\, ,\ 
P_{\OZ|Z}(  1,1)=
P_{\OZ|Z}(  4,4)=1-\xi\, ,
$ 
where $\xi=2\lambda/(1-\lambda)\le 1$. 
Then  $\mu(\rho_{AB})=\ida=I(X;Y|\OZ)=0$ holds, as it should.

If $\lambda>1/3$, then consider the (obviously optimal) channel
$P_{\OZ|Z}(  0,0)=
P_{\OZ|Z}(  2,2)=
P_{\OZ|Z}(  3,3)=
P_{\OZ|Z}(  0,1)=
P_{\OZ|Z}(  0,4)=1
$.
Then
{\small
\[
\mu(\rho_{AB})=\ida=I(X;Y|\OZ)=P_{\OZ}(0)\cdot I(X;Y|\OZ=0)=\frac{1+\lambda}{2}\cdot 
(1-q\log_2 q-(1-q)\log_2(1-q))\ ,
\]
}where $q=2\lambda/(1+\lambda)$.
\exend

  \subsection{Classical Protocols and Quantum Privacy Amplification}\label{prot}

It is a natural question  whether the analogy between entanglement and
intrinsic information (see Section~\ref{linkingone}) carries over to the 
protocol level. The examples given in Section~\ref{exampleszwei} support this 
belief.
A quite interesting and surprising consequence 
 would be that there exists a classical counterpart to bound 
entanglement, namely intrinsic information that cannot be distilled 
into a secret key by any classical protocol, if $|\cx|+|\cy|> 5$.
In other words, the conjecture in~\cite{ittrans}  that such
information can always be distilled would be {\em proved\/} for 
$|\cx|+|\cy|\leq 5$, but {\em disproved\/} otherwise. 

\begin{con}\label{conprot}
Let $\Psi\in\HA\otimes\HB\otimes\HE$ and $\rho_{AB}=\Tr_{\HE}(P_{\Psi})$.
Assume that  for all generating sets  $\{|z\rangle\}$ of $\HE$ there 
are bases $\{|x\rangle\}$ and $\{|y\rangle\}$ of $\HA$ and $\HB$, respectively, 
such that 
$
S(X;Y||Z)>0
$
holds for the distribution $P_{XYZ}(x,y,z):=|\langle x,y,z|\Psi\rangle|^2$.
Then quantum privacy amplification is possible with the state $\rho_{AB}$, 
i.e., $\rho_{AB}$ is free entangled.
\end{con}

\begin{con}
Let $\Psi\in\HA\otimes\HB\otimes\HE$ and $\rho_{AB}=\Tr_{\HE}(P_{\Psi})$.
Assume that  there exists a  generating set $\{|z\rangle\}$ of $\HE$
such that for all 
 bases $\{|x\rangle\}$ and $\{|y\rangle\}$ of $\HA$ and $\HB$, respectively, 
$
S(X;Y||Z)=0
$
holds for the distribution $P_{XYZ}(x,y,z):=|\langle x,y,z|\Psi\rangle|^2$.
Then quantum privacy amplification is impossible with the state $\rho_{AB}$, 
i.e., $\rho_{AB}$ is bound entangled or  separable.
\end{con}

  \subsection{Examples II}
\label{exampleszwei}

The following examples 
support  Conjectures~1 and~2 and illustrate  their
 consequences.
We consider mainly the same distributions as in Section~\ref{exampleseins},
but this time under the aspect of the existence of classical and quantum 
key-agreement protocols. 
\\ \

\noi
{\it Example 1 (cont'd).}
We have shown in Section~\ref{exampleseins} that the 
resulting quantum state is entangled if and only if the intrinsic information
of the corresponding classical situation (with respect to the standard 
bases) is non-zero. Here, we show that such a correspondence also holds on the 
protocol level. First of all, it is clear  for the quantum state that 
QPA is possible whenever the state is entangled because  both $\HA$ and $\HB$
have dimension two.
On the other hand,  the same is also true for 
the corresponding classical situation, i.e.,  secret-key agreement is possible 
whenever
$
\cd/(1-\cd)<2\sqrt{(1-\de)\de}
$
holds, i.e., if the intrinsic information is positive.
This is shown in Appendix~C. There we describe the required protocol, 
more precisely, the advantage-distillation phase (called repeat-code
protocol~\cite{ka}), in which 
Alice and Bob  use
their advantage given by the authenticity of the 
public-discussion channel 
for generating
new random variables for which
the legitimate partners have an advantage over Eve in terms of the (Shannon)
information about each other's new random variables.
For a further discussion of this example, see also~\cite{giswol99}.
\exend
\ \\

\noi
{\it Example 2 (cont'd).}
 The quantum state $\rho_{AB}$ in this example  is bound
entangled, meaning that the entanglement cannot be used for QPA. Interestingly,
but not surprisingly given the discussion above, the corresponding classical 
distribution has the property that $\ida>0$, but nevertheless, all the known 
classical advantage-distillation protocols~\cite{ka},\,  \cite{ittrans}
 fail for this distribution! It seems
that $S(X;Y||Z)=0$ holds (although it is not clear  how this fact could 
be rigorously proven, except by proving Conjecture~1 directly).
\exend
\\ \

\noi
{\it Example 3 (cont'd).}
 We have seen already
that for $2\leq \al\leq 3$, the quantum state is separable and the 
corresponding classical distribution (with respect to the standard bases)
has vanishing intrinsic information. Moreover, it has been shown that for the
quantum situation, $3<\al\leq 4$ corresponds to bound entanglement, 
whereas for $\al>4$, QPA is possible and allows
for generating  a secret key~\cite{dreih}. We describe a classical 
protocol here which suggests
that the situation for the classical translation of the scenario
is totally analogous: The  
protocol allows classical key agreement exactly for $\al>4$. 
However, this does not imply (although it appears very plausible) that 
no classical protocol exists at all for the case $\al\leq 4$.

Let $\al>4$. We consider the following protocol for classical key agreement.
First of all, Alice and Bob both restrict their ranges to $\{1,2\}$ (i.e.,
publicly reject a realization unless $X\in\{1,2\}$ and  $Y\in\{1,2\}$). The
resulting distribution is as follows (to be normalized):
\begin{center}
\begin{tabular}{|c||c|c|}
\hline
\ \ X & 1 & 2\\
Y (Z) &&\\
\hline\hline
1 & (0)\ 2 & (4)\ $5-\al$ \\
\hline
2 & (2)\ $\al$& (0)\ 2 \\
\hline
\end{tabular}
\end{center}
Then, Alice and Bob both send their bits locally over channels $P_{\OX|X}$
and $P_{\OY|Y}$, respectively, such that the resulting bits $\OX$ and 
$\OY$ are symmetric. The channel $P_{\OX|X}$ [$P_{\OY|Y}$] sends $X=0$
[$Y=1$] to $\OX=1$ [$\OY=0$] with probability $(2\al-5)/(2\al+4)$, and
leaves $X$ [$Y$] unchanged otherwise. The  distribution $P_{\OX\OY Z}$ is then
{\small
\begin{center}
\begin{tabular}{|c||c|c|}
\hline
\ \ $\OX$ & 1 & 2\\
$\OY$ (Z) &&\\
\hline\hline
 & (0)\ $2\cdot\frac{9}{2\al+4}$ & (1)\ $5-\al$\\
1 & (2)\ $\al\cdot\frac{9}{2\al+4}\cdot\frac{2\al-5}{2\al+4}$ 
& (2)\ $\al\left(\frac{2\al-5}{2\al+4}\right)^2$ \\
& & (0)\ $2\cdot2\cdot\frac{2\al-5}{2\al+4}$\\
\hline
2 & (2)\ $\al\left(\frac{9}{2\al+4}\right)^2$ & (0)\ $2\cdot \frac{9}{2\al+4}$ \\
&& (2)\ $\al\cdot \frac{9}{2\al+4}\cdot \frac{2\al-5}{2\al+4}$\\
\hline
\end{tabular}
\end{center}
}

It is not difficult to see that for $\al>4$, we have 
$
\prob[\OX= \OY]>1/2
$
and that, given that $\OX=\OY$ holds, Eve has no information at all about
what this bit is. Thus the repeat-code protocol described in Appendix~C
allows for classical key agreement in this situation. 
For $\al\le4$ however, classical key agreement, like quantum key
agreement, seems impossible.
The results of Example~3 are illustrated in Figure~\ref{qcw} in Appendix~A.
\exend
\\ \

\noi
{\it Example 5.}
The following distribution $P_{XYZ}$, with binary $X$ and $Y$, was discussed 
and analyzed in~\cite{ittrans}  as an example of a simple 
distribution for which the equivalence of $\ida>0$ and $S(X;Y||Z)>0$ could
not be shown.

Assume that 
the 
random variables $X$ and $Y$ are distributed 
according to 
$
P_{XY}(0,0)=P_{XY}(1,1)=(1-\al)/2$, $P_{XY}(0,1)=P_{XY}(1,0)=\al/2$,
 and $Z=[Z_X,Z_Y]$, where $Z_X$ and $Z_Y$ 
are generated 
by sending $X$ and $Y$ over 
two independent binary erasure channels with 
erasure probabilities $\de_X$ and $\de_Y$, respectively.

If the conjectured parallels between classical and quantum protocols hold,
then $\ida$ $>0$ implies $S(X;Y||Z)>0$
because both $X$ and $Y$ are binary. Moreover, due to the 
proven connection between intrinsic information and entanglement and hence
 to the eigenvalues of the partial transpose
of the density matrix, the condition for $\ida>0$ can be explicitly 
given, and is very simple. This is  surprising 
since the determination of $\ida$, as well as the advantage-distillation
protocols for this distribution, turned out to be quite 
complicated~\cite{ittrans}.
The condition under which all the eigenvalues of the partial transpose of
the density matrix of the corresponding quantum state are non-negative
is 
\[
(\al-\al^2)\left(\frac{(1-\de_X)(1-\de_Y)}{\de_X}+2\right)
\left(\frac{(1-\de_X)(1-\de_Y)}{\de_Y}+2\right)\geq 1\ .
\]
\cancel{
$
(\al-\al^2)((1-\de_X)(1-\de_Y)/\de_X+2)((1-\de_X)(1-\de_Y)/\de_Y+2)\geq 1
$.
}This bound is compatible with (but stronger than) all the bounds
painfully derived, by working purely in the classical 
information-theoretic world, in~\cite{ittrans}.
\exend

\subsection{Bound Intrinsic Information}\label{secbii}

Examples~2  and~3 suggest that, in analogy to bound entanglement of a quantum state,
{\em bound classical information\/} exists, i.e., conditional intrinsic
information which cannot be used to generate a secret key in the classical
scenario. We give a formal definition of bound intrinsic information.

\begin{defi}
{\rm
Let $P_{XYZ}$ be a  distribution with $\ida>0$. Then if $S(X;Y||Z)>0$
holds for this distribution, the intrinsic information between $X$ and $Y$,
given $Z$, is called {\em free}. Otherwise, if $S(X;Y||Z)=0$, the 
information is called {\em bound}.
}
\defend
\end{defi}

Note that the existence of bound intrinsic information
 could not be proven so far. 
However, all known examples of bound entanglement, combined
with all known advantage-distillation protocols, do not lead
to a contradiction to Conjecture~1! Clearly, it would be 
very interesting to rigorously prove this conjecture because then, all
pessimistic results  known for the quantum scenario would 
immediately carry over to the classical setting  (where such results 
appear to be much harder to prove).

Examples~2 and~3 also illustrate nicely what the nature of such bound 
information is. Of course, $\ida>0$ implies both $I(X;Y)>0$ and 
$I(X;Y|Z)>0$. However, if $|\cx|+|\cy|>5$, it is possible that 
the dependence between $X$ and $Y$ and the dependence between 
$X$ and $Y$, given 
$\OZ$, are ``orthogonal.'' By the latter we 
mean that for all fixed (deterministic or probabilistic) functions 
$f\, :\, \cx\rightarrow \{0,1\}$ and $g\, :\, \cy\rightarrow \{0,1\}$ for which
the correlation of $f(X)$ and $g(Y)$ is positive, i.e., 
\[
P_{f(X)g(Y)}(0,0)\cdot P_{f(X)g(Y)}(1,1)>
P_{f(X)g(Y)}(0,1)\cdot P_{f(X)g(Y)}(1,0)\ ,
\]
the correlation between the same binary random variables,
given $\OZ=\oz$, is negative (or ``zero'') for  all
 $\overline{z}\in\overline{{\cal Z}}$, where  $\OZ$ is the 
random variable generated by sending $Z$ over Eve's optimal channel
$P_{\OZ|Z}$.

A complete understanding of bound intrinsic information 
is of interest also because it 
automatically leads to 
a better understanding of bound entanglement in quantum information
theory.

\section{Concluding Remarks}
We have considered the model of information-theoretic key agreement 
by public discussion from correlated information. More precisely, 
we have compared  scenarios where the joint information is 
given by classical random variables and by quantum states (e.g., after 
execution of a quantum protocol). We proved a close connection between
such classical and quantum information,  namely between
intrinsic information and entanglement.
As an application, the derived parallels lead to an efficiently verifiable 
criterion for the fact that the intrinsic information vanishes. Previously, 
this quantity was considered to be quite hard to handle.

Furthermore,
we have presented examples 
providing
 evidence for the fact that the close connections between classical and quantum 
information extend to the level of the 
protocols. As a consequence, the powerful tools and statements 
on the existence or rather non-existence of quantum-privacy-amplification
protocols immediately carry over to the classical scenario, where it 
is often  unclear how 
to show that no protocol exists.
In particular, many examples
(only some of which are presented above due to space limitations)
coming from measuring bound entangled states, and for which none 
of the known classical secret-key agreement protocols is successful,
strongly
suggest that bound entanglement  has a classical counterpart:
intrinsic information which cannot be distilled to a secret key.
This stands in sharp contrast to what was previously believed about 
classical key agreement.
We state as an open problem to rigorously prove Conjectures~1 and~2.

Finally, we have proposed a measure for entanglement, based on classical 
information theory, with all  the properties required for such a measure.

\newpage
\section*{Appendix A: Figures}

\begin{figure}[h]
\hbox{
\centerline{
\psfig{figure=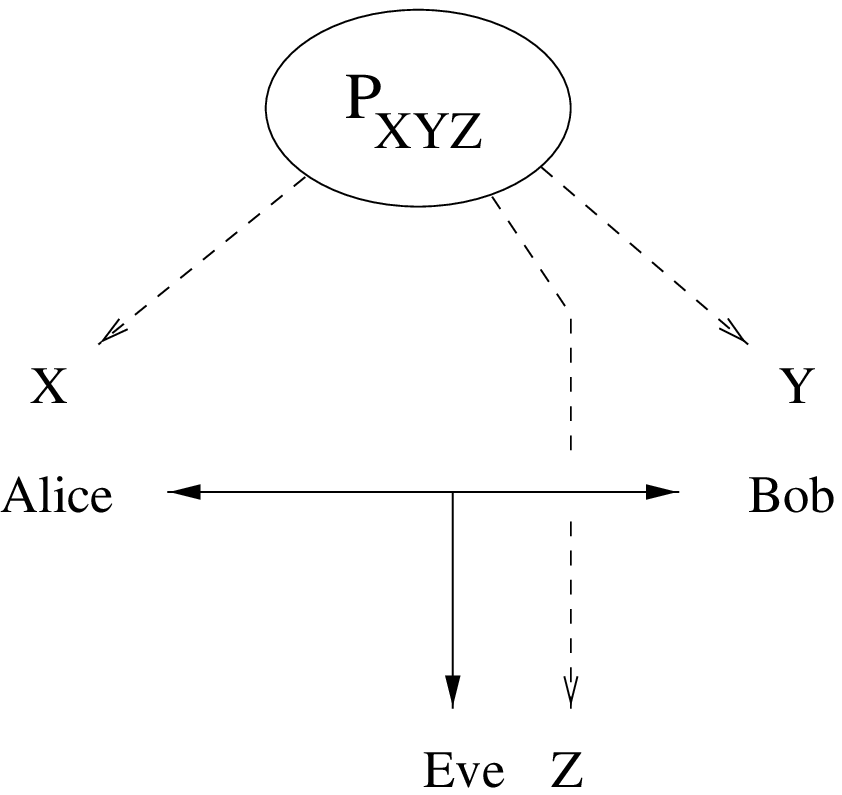,width=6cm}}}
\caption{Secret-Key Agreement from Common Information}
\label{maumod}
\end{figure}

\begin{figure}[h]
\hbox{
\centerline{
\psfig{figure=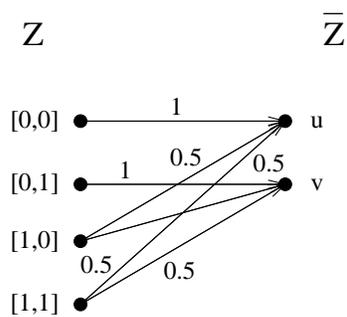,width=4.5cm}}}
\caption{The Channel $P_{\OZ|Z}$ in Example~1}
\label{channel}
\end{figure}

\nopagebreak

\begin{figure}[!h]
\hbox{
\centerline{
\psfig{figure=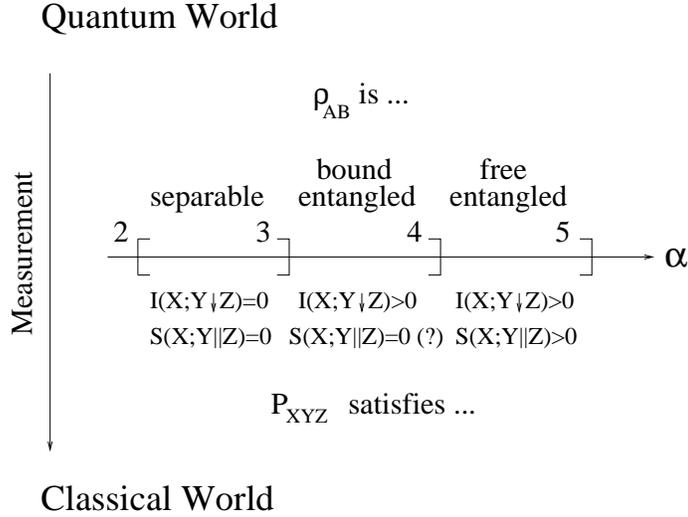,width=9cm}}}
\caption{The Results of Example 3}
\label{qcw}
\end{figure}

\newpage
\section*{Appendix B: Measuring in ``bad'' Bases}

In this appendix we show, by two examples, that the statements of Theorems~1 and~2
do not always hold for the standard bases and, in particular, not for 
arbitrary bases: Alice and Bob as well as Eve can perform bad measurements 
and give away an initial advantage. Let us begin with an example where 
measuring in the standard basis is a bad choice for Eve.
\\ \

\noi
{\it Example 6.}
Let us consider the quantum states
\[
\Psi  =  \frac{1}{\sqrt{5}}\, (|00+01+10\rangle \otimes|0\rangle  + |00+11\rangle \otimes|1\rangle )\ ,\ \ 
\rho_{AB}  =  \frac{3}{5}\,  P_{|00+01+10\rangle } + \frac{2}{5}\,  P_{|00+11\rangle }\ .
\]
If Alice, Bob, and Eve measure in the standard bases, we get the classical distribution 
 (to be normalized)
\begin{center}
\begin{tabular}{|c||c|c|}
\hline
\ \ X & 0 & 1\\
Y (Z) &&\\
\hline\hline
0 & (0)\ 1 & (0)\ 1 \\
 &  (1)\ 1 & (1)\ 0 \\
\hline
1 & (0)\ 1 & (0)\ 0 \\
  & (1)\ 0 & (1)\ 1\\
\hline
\end{tabular}
\end{center}
For this distribution, $\ida>0$ holds. Indeed, 
even $S(X;Y||Z)>0$ holds.
This is not surprising since both $X$ and $Y$ 
are binary, and since the described parallels suggest that in this 
case, positive intrinsic information implies that a secret-key 
agreement protocol exists. 

The proof of $S(X;Y||Z)>0$ in this situation is analogous to the proof of this 
fact in Example~3. The protocol consists of Alice and Bob independently 
making their bits symmetric. Then the repeat-code protocol can be 
applied. 

However, the 
partial-transpose condition shows that $\rho_{AB}$ is separable. This 
means that measuring in the standard basis is bad for Eve. Indeed,
let us rewrite $\Psi$ and $\rho_{AB}$ as
\[
\Psi  =  \sqrt{\Lambda}\, |\vec m,\vec m\rangle \otimes|\tilde 0\rangle + 
     \sqrt{1-\Lambda}\, |-\vec m,-\vec m\rangle \otimes|\tilde 1\rangle\ ,\ \ 
\rho_{AB}  =  \frac{5+\sqrt{5}}{10}\,  P_{|\vec m,\vec m\rangle } + 
         \frac{5-\sqrt{5}}{10}\,  P_{|-\vec m,-\vec m\rangle }\ ,
\]
where
$\Lambda  =  (5+\sqrt{5})/10$, 
$|\vec m,\vec m\rangle  =  |\vec m\rangle \otimes|\vec m\rangle$,
$|\pm\vec m\rangle  =  \sqrt{(1\pm\eta)/2}\, |0\rangle  \pm \sqrt{(1\mp\eta)/2}\, |1\rangle$, and 
$\eta  =  1/\sqrt{5}$.

In this representation, $\rho_{AB}$ is obviously separable. It also means that 
 Eve's optimal measurement basis is
\[
|\tilde 0\rangle =\sqrt{\Lambda}\, |0\rangle  - \frac{1}{\sqrt{5\Lambda}}\, |1\rangle \ ,\ \ 
|\tilde 1\rangle =-\sqrt{1-\Lambda}\, |0\rangle  - \frac{1}{\sqrt{5(1-\Lambda)}}\, |1\rangle\ . 
\]
Then, $\ida=0$ holds for the resulting classical distribution.
\exend
\ \\

\noi
Not surprisingly, there also exist examples of distributions for which measuring 
in the standard bases is bad for Alice and Bob. These states are entangled, but $\ida=0$
holds.
\\ \

\noi
{\it Example 7.}
Let the following classical distribution be given:
\begin{center}
\begin{tabular}{|c||c|c|}
\hline
\ \ X & 0 & 1\\
Y (Z) &&\\
\hline\hline
0 & (0)\ 0.0082 & (0)\ 0.0219 \\
 &  (1)\ 0.0006 & (1)\ 0.0202 \\
\hline
1 & (0)\ 0.0729 & (0)\ 0.3928 \\
  & (1)\ 0.0905 & (1)\ 0.3889204545\\
\hline
\end{tabular}
\end{center}

Because of 
\[
(0.0082+0.0006)\cdot (0.03928+0.3889204545)=(0.0219+0.0202)\cdot (0.0729+0.0905)
\]
we have $I(X;Y)=0$, thus $\ida=0$.
On the other hand, the corresponding quantum state,
for which the above distribution results by measuring in the standard bases,
can be shown to be entangled. 
\exend

\section*{Appendix C: A Protocol for Advantage Distillation}

 The following protocol for classical advantage distillation is called 
  {\em repeat-code protocol\/} and was first proposed in~\cite{ka}. 
Note that there exist  more 
efficient protocols in terms of the amount of extractable secret key.
However, since we only want to prove  qualitative possibility results,
 it is sufficient to look at this simpler protocol.
Assume the scenario of Example~1.

Let $N>0$ be an even integer, and let Alice choose a random bit $C$ and send
the block
\[
X^N\oplus C^N:=[X_1\oplus C,X_2\oplus C,\ldots, X_N\oplus C]
\]
over the classical channel.
Here, $X^N$ stands for the block $[X_1,X_2,\ldots, X_N]$ of $N$ consecutive
realizations of the random variable $X$,
whereas $C^N$ stands for the $N$-bit block $[C,C,\ldots,C]$.
Bob then computes
$[(C\op X_1)\op Y_1,\ldots,(C\op X_N)\op Y_N]$ and (publicly)
accepts exactly if this block
is equal to either $[0,0,\ldots,0]$ or $[1,1,\ldots,1]$. In other words,
Alice and Bob make use of a repeat code of length $N$ with the only two
 codewords
$[0,0,\ldots,0]$ and $[1,1,\ldots,1]$.

Bob's conditional error
probability $\be_N$
when guessing the bit sent by Alice, given that he accepts, is
\[
\be_N=\frac{1}{p_{a,N}}\cdot D^N\leq\left(\frac{D}{1-D}\right)^N\ ,
\]
where $p_{a,N}=D^N+(1-D)^N$ 
is the probability that Bob accepts the received 
block.
It is obvious that Eve's optimal 
strategy for guessing $C$  is to compute the block
$[(C\oplus X_1)\oplus Z_1,\ldots,(C\oplus X_N)\oplus Z_N]$ and guess $C$ as
$0$ if at least half of the bits in this block are $0$, and as $1$ otherwise.
Given that Bob correctly accepts, Eve's error probability when guessing the
bit $C$ with the optimal strategy
is lower bounded by $1/2$ times the probability that she decodes
to a block with $N/2$ bits $0$ and the same number of $1$'s. Hence
we get that 
\[
\ga_N\geq \frac{1}{2}{N \choose N/2}(1-\de)^{N/2}\de^{N/2}
\geq\frac{K}{\sqrt{N}}\cdot\left(2\sqrt{(1-\de)\de}\right)^N
\] 
holds for some constant $K$ and 
for sufficiently large $N$ by using Stirling's formula.
Note that Eve's error probability given that Bob accepts is 
asymptotically equal to her error probability given that 
Bob {\em correctly\/} accepts because Bob accepts erroneously
only with asymptotically vanishing probability, given that he 
accepts.

Although it is not the adversary's ultimate goal to guess the bits 
$C$ sent by Alice, it has been shown 
that the fact that $\be_N$ decreases exponentially faster than $\ga_N$
implies that for sufficiently large $N$, Bob has more (Shannon) information
about the bit $C$ than Eve (see for example~\cite{ittrans}). 
Hence Alice and Bob have managed to generate
new random variables with the property that Bob obtains more information about 
Alice's random bit than Eve has. Thus $S(X;Y||Z)>0$ holds.

\end{document}